\documentclass[%
 reprint,
 amsmath,amssymb,
pra,
showkeys,
]{revtex4-2}

\usepackage{graphicx}
\usepackage{dcolumn}

\usepackage{bm}
\usepackage{gensymb}
\usepackage{mathrsfs}

\usepackage{titlesec}
\usepackage{color}
\usepackage{tikz}

 \titlespacing{\section}{-6pc}{3.5ex plus .1ex minus .2ex}{1.5ex minus .1ex}

\usepackage{tcolorbox}
\usepackage{soul}

\usepackage{enumitem}
\newlist{steps}{enumerate}{1}
\setlist[steps, 1]{label = Step \arabic*:}

\usepackage[toc,page]{appendix}

\usepackage{xcolor}
\usepackage{url}

\newcommand{\red}[1]{\textcolor[rgb]{1.0,0.0,0.0}{#1}}

\newcommand{\black}[1]{\textcolor[rgb]{0.0,0.0,0.0}{#1}}

\newcommand{\green}[1]{\textcolor[rgb]{0.0,1.0,0.0}{#1}}

\newcommand{\postb}[1]{\textcolor[rgb]{0.30,0.75,0.93}{#1}}

\usepackage{scalerel}
\usepackage{tikz}
\usetikzlibrary{svg.path}
\definecolor{orcidlogocol}{HTML}{A6CE39}
\tikzset{
  orcidlogo/.pic={
    \fill[orcidlogocol] svg{M256,128c0,70.7-57.3,128-128,128C57.3,256,0,198.7,0,128C0,57.3,57.3,0,128,0C198.7,0,256,57.3,256,128z};
    \fill[white] svg{M86.3,186.2H70.9V79.1h15.4v48.4V186.2z}
                 svg{M108.9,79.1h41.6c39.6,0,57,28.3,57,53.6c0,27.5-21.5,53.6-56.8,53.6h-41.8V79.1z M124.3,172.4h24.5c34.9,0,42.9-26.5,42.9-39.7c0-21.5-13.7-39.7-43.7-39.7h-23.7V172.4z}
                 svg{M88.7,56.8c0,5.5-4.5,10.1-10.1,10.1c-5.6,0-10.1-4.6-10.1-10.1c0-5.6,4.5-10.1,10.1-10.1C84.2,46.7,88.7,51.3,88.7,56.8z};
  }
}
\newcommand\orcidicon[1]{\href{https://orcid.org/#1}{\mbox{\scalerel*{
\begin{tikzpicture}[yscale=-1,transform shape]
\pic{orcidlogo};
\end{tikzpicture}
}{|}}}}
\usepackage{hyperref}       

\begin{document}

\title{An Interpretable Machine Learning Model  for Deformation of Multi--Walled Carbon Nanotubes}

\author{Upendra Yadav \orcidicon{0000-0002-4863-3929}, Shashank Pathrudkar \orcidicon{0000-0001-8546-8056}, and  Susanta Ghosh \orcidicon{0000-0002-6262-4121}}
\email[Address for correspondence: ]{susantag@mtu.edu}

\affiliation{Department of Mechanical Engineering--Engineering Mechanics and The Center for Data Sciences, Michigan Technological University, MI, USA}

\newcommand \blackline {\raisebox{2pt}{\tikz{\draw[-,black!40!black,solid,line width=1pt](0,0)--(5mm,0);}}}
\newcommand \redline {\raisebox{2pt}{\tikz{\draw[-,red,dashed,line width = 1pt](0,0) -- (5mm,0);}}}
\newcommand \greenline {\raisebox{2pt}{\tikz{\draw[-,green,solid,line width = 1pt](0,0) -- (5mm,0);}}}
\newcommand \blueline {\raisebox{2pt}{\tikz{\draw[-,blue,solid,line width = 1pt](0,0) -- (5mm,0);}}}
\definecolor{post_b}{rgb}{0.30,0.75,0.93}
\newcommand \redcircle {\raisebox{0pt}{\tikz{\draw[red,fill=none] (0,0) circle (.5ex);}}}

\begin{abstract} 
We present a novel interpretable machine learning model to accurately predict complex rippling deformations of Multi–Walled Carbon Nanotubes (MWCNTs) made of millions of atoms. Atomistic–physics–based models are accurate but computationally prohibitive for such large systems. To overcome this bottleneck, we have developed a machine learning model that consists of a \black{novel} dimensionality reduction technique and a deep neural network–based learning in the reduced dimension. The proposed nonlinear dimensionality reduction technique \black{extends the functional principal component analysis to satisfy the constraint of deformation. Its novelty lies in designing a function space that satisfies the constraint exactly, which  is crucial for efficient dimensionality  reduction}. 
Owing to the dimensionality reduction and several other strategies adopted in the present work, learning through deep neural networks is remarkably accurate. The proposed model accurately matches an atomistic–physics–based model while being orders of magnitude faster. It extracts universally dominant patterns of deformation in an unsupervised manner. These patterns are comprehensible and elucidate how the model predicts, yielding interpretability. The proposed model can form a basis for the exploration of machine learning toward the mechanics of one and two--dimensional materials.  
\end{abstract}

\maketitle

\section{\label{sec:Introduction}INTRODUCTION}
Carbon nanotubes have shown remarkable physical, chemical and electronic properties. Moreover, their deformation can be used to control their chemical, and electronic properties, leading to a large number of applications including nanoelectromechanical systems \cite{torsional_NEMS,papadakis04,DeVolder2013Science}. 
In experiments, multi--walled carbon nanotubes (MWCNTs) show periodic wavelike deformation patterns called rippling \cite{Poncharal1999Science,papadakis04}. 
Accurate and efficient simulation tools to predict the complex deformations of large MWCNTs are needed but still elusive. 
Quantum--mechanical and molecular simulations are accurate but they are computationally prohibitive for large MWCNTs containing millions of atoms. 
Towards this, Atomistic--Continuum (AC) models have been developed \cite{Arroyo2002JMPS,Arroyo2004IJNME,Ghosh} by integrating atomistic and continuum frameworks. 
State--of--the--art AC models are efficient but still require a significant amount of high--performance computing efforts for large MWCNTs, which is the bottleneck for exploration of the physics of these materials. 

Machine Learning (ML) methods such as Deep Neural Networks (DNNs) \cite{Lecun2015,HORNIK1989359} are intensely investigated for accelerating mechanics, physics, and materials research \cite{Xie2018PRL,Iten2020PRL,Lu7052}, however, so far most of the applications are limited to the prediction of  low--dimensional properties, such as material moduli. On the contrary, discretized material deformation requires prediction in a high--dimensional space.
For instance, large thick MWCNTs  require several millions of degrees of freedom to describe its deformation \cite{Arias2008PRL,Ghosh}. 

\black{Deep Learning models can predict low dimensional (e.g. CNN, Autoencoder}\cite{hanakata2018prl, hanakata2020arxiv}\black{) or high dimensional outputs (e.g. Encoder-Decoder}  \cite{gomez2018acs, balu2019sr}\black{). However, these Deep Learning models require high dimensional inputs.} State--of--the--art {DNNs cannot  accurately predict high--dimensional targets \black{from a few input features}.}
The objective of the present study is to create an ML model to accurately and efficiently predict high--dimensional  discretized deformations of MWCNTs \black{ as output from  low--dimensional inputs. This calls for a dimensionality reduction for the outputs.}

\begin{figure*}[htbp]
    \centering
    \includegraphics[width=1\linewidth]{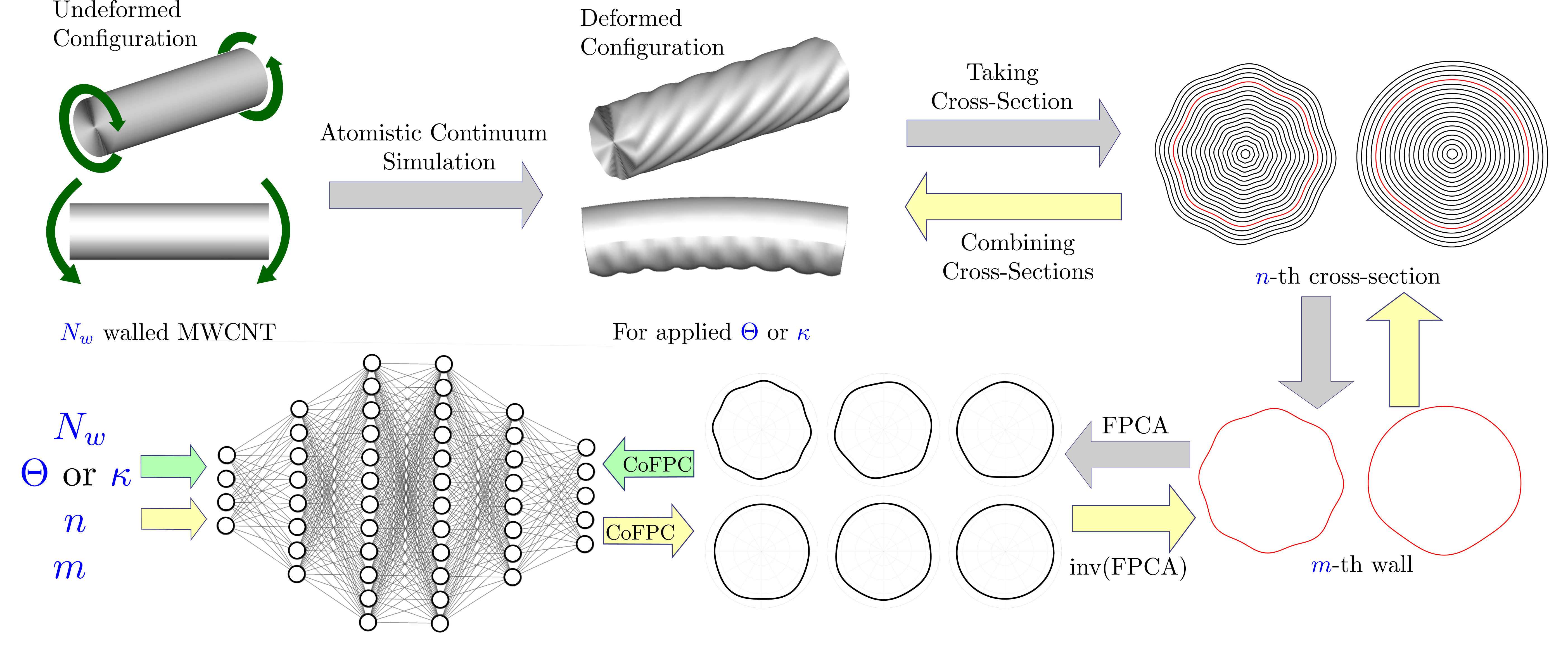}    
    \caption{Schematic of the present framework involving the data generation via AC simulation and the proposed Deformation Manifold Learning(DML) model. The DML model includes: c--FPCA and DNN. Firm arrows show data generation and training, and dashed arrows show prediction via the proposed model.}
    \label{fig:DLM_schematic}
\end{figure*}

An additional challenge for the deformation of MWCNTs  is that the manifold of reduced--dimension  (called \emph{latent space}) is non--linear. Nonlinear Dimensionality Reduction techniques (also called Manifold Learning) such as Isomap,  Locally--Linear Embedding, and Umap are designed to identify the low--dimensional non--linear manifold structure of the data  \cite{Book_NonlinearDR_Lee_Verleysen,mcinnes2018umap}. 

However, to accurately predict the deformation of MWCNTs we need an accurate, smooth, and functional representation of the mapping from the high--dimensional to a low--dimensional manifold such that it respects the constraints of the deformation. 
Just visualization or approximate discrete representations of the low--dimensional manifold are not sufficient for the present purpose.

Functional Principal Component Analysis (FPCA)  \cite{yao2005,ramsayfunctional} provides a smooth functional representation of the data,  
which is analogous to Kosambi--Karhunen--Loève Expansion  \cite{Stark1986Book,Jorgensen2007}. 
FPCA  represents a stochastic process through a linear combination of an infinite number of orthogonal functions. 
\black{However, FPCA cannot respect the geometric constraint since these orthogonal functions need not satisfy any constraint, as a consequence FPCA yields discontinuous and erroneous predictions for MWCNTs  that has periodicity constraint along the circumference.} 
\black{In this work, we propose to extend FPCA by designing a basis set of functions to satisfy this constraint exactly. We call the proposed technique constrained--FPCA (c--FPCA)}. 
The proposed \black{c--FPCA} technique alleviates the \emph{curse of dimensionality} by providing low--dimensional functional representations for the deformations of MWCNTs.

The proposed  semi--supervised ML model includes two steps (i) unsupervised  dimensionality reduction (via proposed c--FPCA) of the deformed manifold and (ii) supervised learning (via DNN) of deformation in the reduced dimension. Henceforth, the proposed ML model is referred to as the  \emph{Deformation Manifold Learning} (DML) model, shown in Fig.~\ref{fig:DLM_schematic}. It takes the details of the MWCNT system and its boundary conditions as inputs and predicts its high--dimensional discretized deformation. 
\section{DEFORMATION MANIFOLD LEARNING MODEL} \label{sec:DLM}
We focus on complex distributed periodic buckling patterns (rippling) of MWCNTs under torsion and bending \cite{Arroyo2003PRL}. 
An atomistic--continuum model (namely \emph{Foliation} model \cite{Ghosh}) is used to generate the training data for the proposed DML model. To represent thick MWCNTs commonly found in the experiments, we are simulating $(5,5),(10,10),\cdots, (5\,N_{w}, 5\,N_{w})$ MWCNTs walls, with $N_w = 10$ to $N_w = 40$ in the increment of 5, where $N_{w}$ is the number of walls. 
\black{These 7 simulations ($N_w = 10,15,\cdots,40$) are used in training.}

\begin{figure}[ht]
    \includegraphics[width=1.0\linewidth]{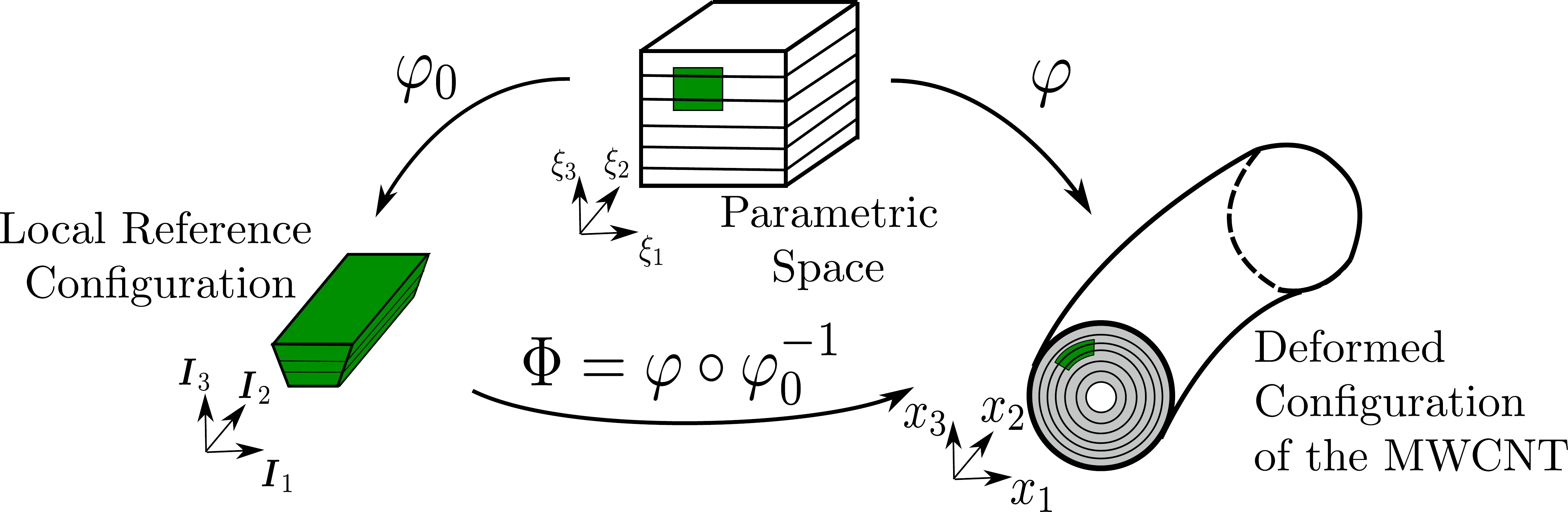}
    \caption{Illustration of the kinematics showing the configuration spaces, deformation maps, and coordinate axes.}
    \label{fig:fol}
\end{figure} 

\subsection{Kinematics of MWCNTs and Data Preparation}\label{sec:DataPreprocessing}
A configuration map ($\varphi_0$) is an injective mapping from parametric space to the local reference configuration. Another map ($\varphi$) is from parametric space to Euclidean space, $\mathbb{R}^3$ as shown in Fig.~\ref{fig:fol}. The deformation map is then defined as $\Phi = \varphi \circ \varphi_0^{-1}$. 

To learn the deformation pattern of entire MWCNT, multiple sets of simulations would be required.
To reduce the number of simulations we decomposed the domain into several cross--sections (\black{$N_{cs}$}) 
at regular intervals along its length. Due to the periodicity of the rippling deformation\black{,} this decomposition strategy increases the size of the data set. 
\black{Following the decomposition technique, } the total deformation \black{of each point} \black{($\xi_1,\xi_2,\xi_3$)} for $m$--th wall at $n$--th  cross--section of a MWCNT can be represented through two parts, (i) an in--plane radial deformation  $r(\theta(\xi_1),\xi_2^n,\xi_3^m)$ in the undeformed cross--sectional plane,  and (ii) axial deformation   $\Phi_2((\theta(\xi_1),\xi_2^n,\xi_3^m))$. 

\subsection{Dimensionality Reduction through  \black{Proposed Constrained}--FPCA}\label{append:FPCA} 
The cross--sections of the MWCNTs are given by the mapped $(\xi_1,\xi_3)$ planes for different $\xi_2$ in the deformed configuration, which constitutes the data set  $\left\{r_i(\theta)\right\}_{i=1}^{N}$ of length $N$. 
Let us assume the radial deformations of each tube are sampled from a stochastic process $R(\theta)$, $\theta \in \mathcal{T}=(0,2\pi)$, such that its second derivative is square--integrable. This smoothness of the deformation map is a necessary condition since the energy of a MWCNT is a function of curvature of its walls.  We suppose that $R(\theta)$ can take any of the values $r_i(\theta)\, \in \mathscr{H}^{2}(\mathcal{T}),\, i = 1,\cdots,N$, such that $g(r_i(\theta))=0$. Where $g(r_i(\theta))=0$, is a geometric constraint on the  deformation of MWCNTs. Where $\mathscr{H}^{2}(\mathcal{T})$ is Hilbert space. 
We denote the $L^2(\mathcal{T})$ inner product of functions   $\phi_i\,,\, \phi_j \in \mathscr{H}^{2}(\mathcal{T})$  with $\langle \phi_i, \phi_j \rangle : \int_{\mathcal{T}} \phi_i(\theta)\,\phi_j(\theta)\,d\theta$.

Let the mean and the covariance functions of $R(\theta)$ are denoted by $\mu(\theta)$ and $v(\theta,\vartheta) = \mathrm{Cov}(R(\theta),R(\vartheta))$. Invoking the  also known as the Kosambi--Karhunen--Loève Expansion theorem \cite{Stark1986Book,Jorgensen2007}, the centered process can be expressed as 
$R(\theta)-\mu(\theta) = \sum_{k=1}^{\infty}\,\bar{c}_k\,\psi_{k}(\theta)$.
Here, $\bar{c}_k = \langle (R(\theta)-\mu(\theta)),\psi(\theta) \rangle$. 
Where  $\psi_{k}(\theta),\, k=1,2,\cdots,$ are the orthonormal eigenfunctions of the following eigenvalue problem  ${\int_{\mathcal{T}} v(\theta,\vartheta)\,\psi(\theta)d\vartheta = \lambda\,\psi(\theta)}$.

These eigenfunctions, $\psi_k(\theta)$, are henceforth   referred to as  functional principal components (functional--PCs).
Assuming a finite set of eigenfunctions is sufficient to  approximate the centered stochastic process, $R(\theta)-\mu(\theta)$,   its $i$--th sample can be written as 
\begin{eqnarray}\label{eqn:truncatedKLE}
    {r}_i(\theta) - \mu(\theta) \approx \sum_{k=1}^{K}\, \bar{c}_{ik}\, \psi_{k}(\theta), \; i=1, \cdots, N
\end{eqnarray}
Interpretation of eigenfunctions:
The $k$--th eigenfunction  
$\psi_{k}$ is the $k$--th most dominant mode of variation orthogonal to $\left\{\psi_{i}\right\}_{i=1}^{k-1}$. 

Solving an eigenvalue problem in $ \mathscr{H}^{2}(\mathcal{T})$ while satisfying a constraint can be a difficult task, therefore, we choose a convenient finite--dimensional basis and look for solutions in terms of that predefined basis. However, choosing any arbitrary basis will not work, since the deformation configurations of MWCNTs have  geometric constraints that need to  be satisfied by the eigenfunctions and hence also needs to  be satisfied by the basis. \black{Since FPCA with an arbitrary basis does not satisfy this constraint it does not work for MWCNT.} 
We choose a basis $\mathcal{B} = \left\{\phi_{k} \in \mathscr{H}^{2}(\mathcal{T}),\, g(\phi_{k})=0,\, k=1,\cdots,K \right\}$. 
 We encode the geometric constraint \black{(periodicity constraint)} of the  deformation of MWCNTs via the function $g(\phi_{k})=0$, which is crucially important and specializes the FPCA for the present system. \black{We call this novel technique constrained--FPCA (c--FPCA)}.

We rewrite the data set, $\left\{r_i(\theta)\right\}_{i=1}^{N}$, 
the eigenfunction $\psi(\vartheta)$, and the covariance function $v(\vartheta,\theta)$ in terms of the basis $\mathcal{B}$ and solve the aforementioned eigenvalue problem 
to obtain the functional--PCs, $\psi_k(\vartheta)$. Subsequently, the function $r_i(\theta)$ is represented in terms of  functional--PCs using the Eq.~\ref{eqn:truncatedKLE} and their corresponding coefficients ($\bar{c}_{ik}$) are referred here as coefficients of functional--PCs \mbox{(CoFPCs)}. The dimension of the problem is significantly reduced by obtaining a $K$ (number of functional--PCs) much smaller than the  size of the discretized ${r}_i(\theta)$. 

\subsection{Learning in the  Reduced  Dimension  through Deep Neural  Networks}\label{sec:DNN}
 
We have developed Deep Neural Networks (DNNs)  to map the MWCNT system parameters to its deformation in the reduced dimension.
The DNN architecture takes the Geometry parameters and Boundary conditions as input and outputs CoFPCs. 
\black {The  4 Inputs for the proposed DNN are: Geometry parameters (i) total number of walls in the MWCNT ($N_{w}$), (ii) the wall number ($m, m = 1,\cdots, N_{w}$), and (iii) the length coordinate ($\Phi_2(\xi_2^n), n = 1,\cdots, N_{\mathrm{cs}}$); (iv)  Boundary Conditions: Angle of twist ($\Theta$) or  Curvature ($\kappa$), per unit length.} 
\black{The dimension of the output layer is the number of CoFPCs, which is decided based on the accuracy required (in c--FPCA), details of which are provided in Sec.} \ref{sec:Results} \black{A.}
The DNN architecture   consists of approximately 40 thousand learning parameters \black{ and uses mean squared error of CoFPCs as the  loss function}.

DNNs are prone to overfitting and often fail to work accurately for test data.
To alleviate the overfitting of the DNN  multiple regularization \cite{Prechelt2012,zou2005regularization} and normalization \cite{jayalakshmi2011statistical} strategies  are  adopted in the present work. 

Three DNNs are trained for predicting the following deformations of MWCNTs: (i) In--plane  deformation under torsion, (ii) in--plane, and (ii) out--of--plane (axial) deformation under bending. Unlike torsion, in bending the axial deformation is not negligible, hence we have used two  DNNs for in--plane and axial deformations.

\section{Results }\label{sec:Results} 

\subsection{Dimensionality Reduction}
\begin{figure}[htbp]
    \centering
    \includegraphics[width=1\linewidth]{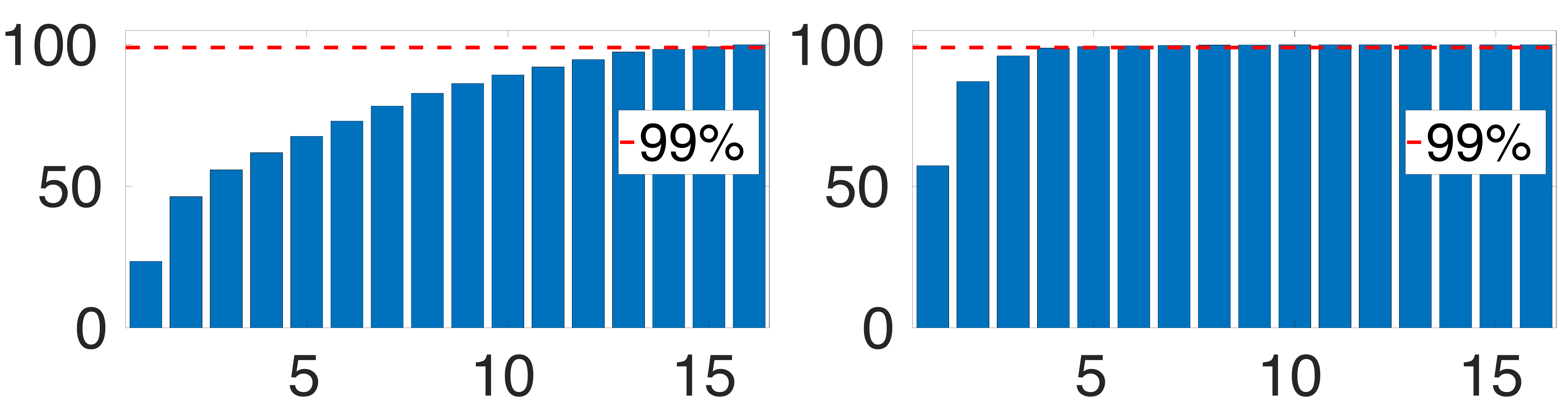}
    \caption{Cumulative \% variance captured by principal components  for MWCNTs under  torsion (left) and bending (right).}
    \label{fig:fPCA_DR}
\end{figure}

The proposed dimensionality reduction could  capture $99\%$ variability of the deformation data set through only 14 and 4 functional--PCs for torsion and bending respectively, as shown in Fig.~\ref{fig:fPCA_DR}. To capture $99.9$\% variability, the corresponding numbers are 16 and 6 respectively. \black{The associated (16 and 6) CoFPCs are used as the outputs of DNNs.}
To obtain the functional--PCs we started with 64 basis functions to represent data vectors of size up to several hundred. 
This demonstrates up to two orders of magnitude dimensionality reduction via the present approach. \black{Owing to the high accuracy of c--FPCA,  DNNs need to learn in  significantly reduced dimensions, yielding higher accuracy. Further, c--FPCA returns only a few  dominant modes having a  perspicuous pattern, which makes it easier for DNN to learn the pattern.}

\begin{figure}[h]
    \includegraphics[width=1\linewidth]{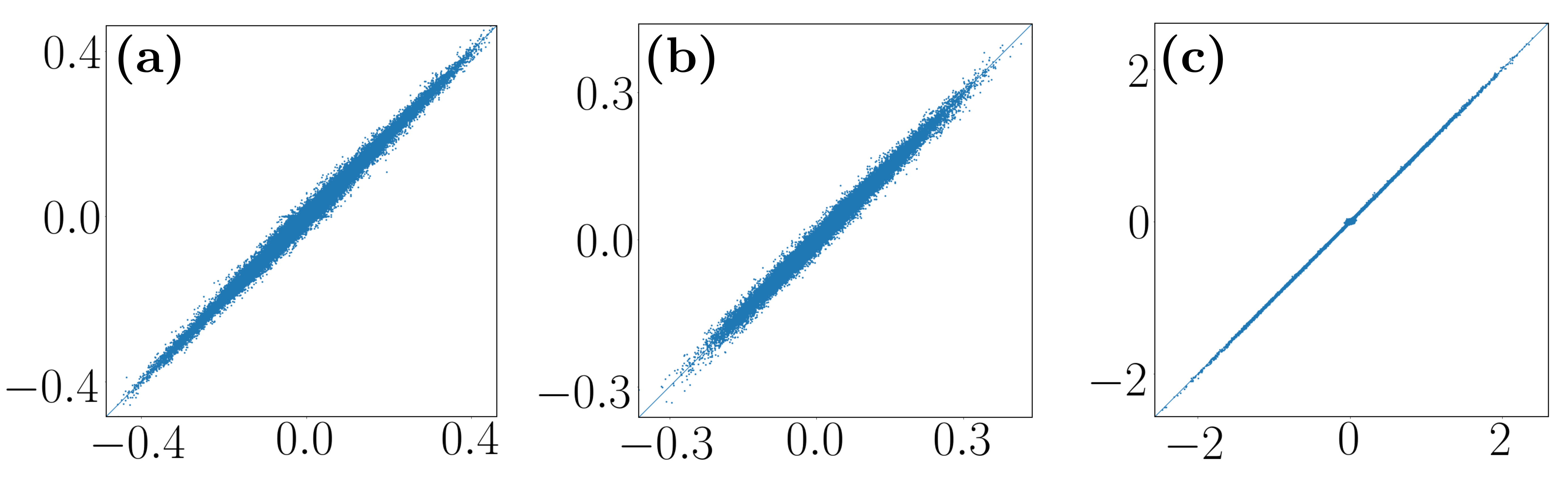}
 
    \caption{Correlation plots for \black{test set CoFPCS of} (a) in--plane deformation in torsion, (b) in--plane and (c) out--of--plane deformation in bending.  \textit{R} = $0.9943$(a), $0.9931$(b), $0.9991$(c).}  
    \label{fig:coorelation_torsion_bending}
\end{figure}

\subsection{Accuracy}
\begin{figure}[t]
    \centering
     \includegraphics[width=1\linewidth]{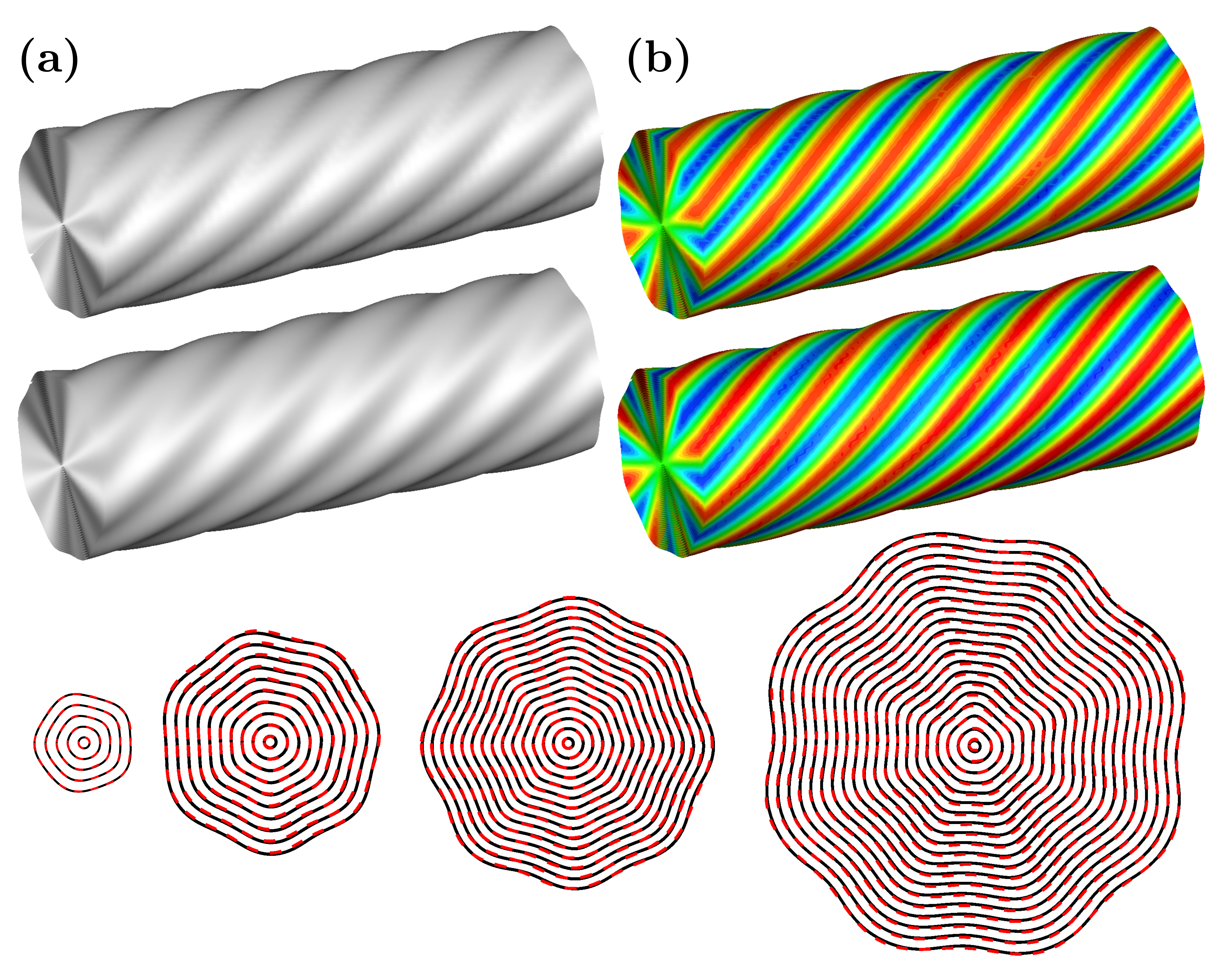}

    \caption{ (a)  Twisted 40 walled CNT obtained via AC (top) and DML (bottom) model. (b) Radial deformation colormap (Red: high, Blue: low). Alternate walls of cross--sections obtained via AC ({\black{$\boldsymbol{-}$}}) and DML ({\red{$\boldsymbol{- -}$}}) models, for 10, 20, 30, and 40 walled CNTs.}

    \label{fig:torsion_accuracy}
\end{figure}

\begin{figure}[t]
    \centering
    \includegraphics[width=1\linewidth]{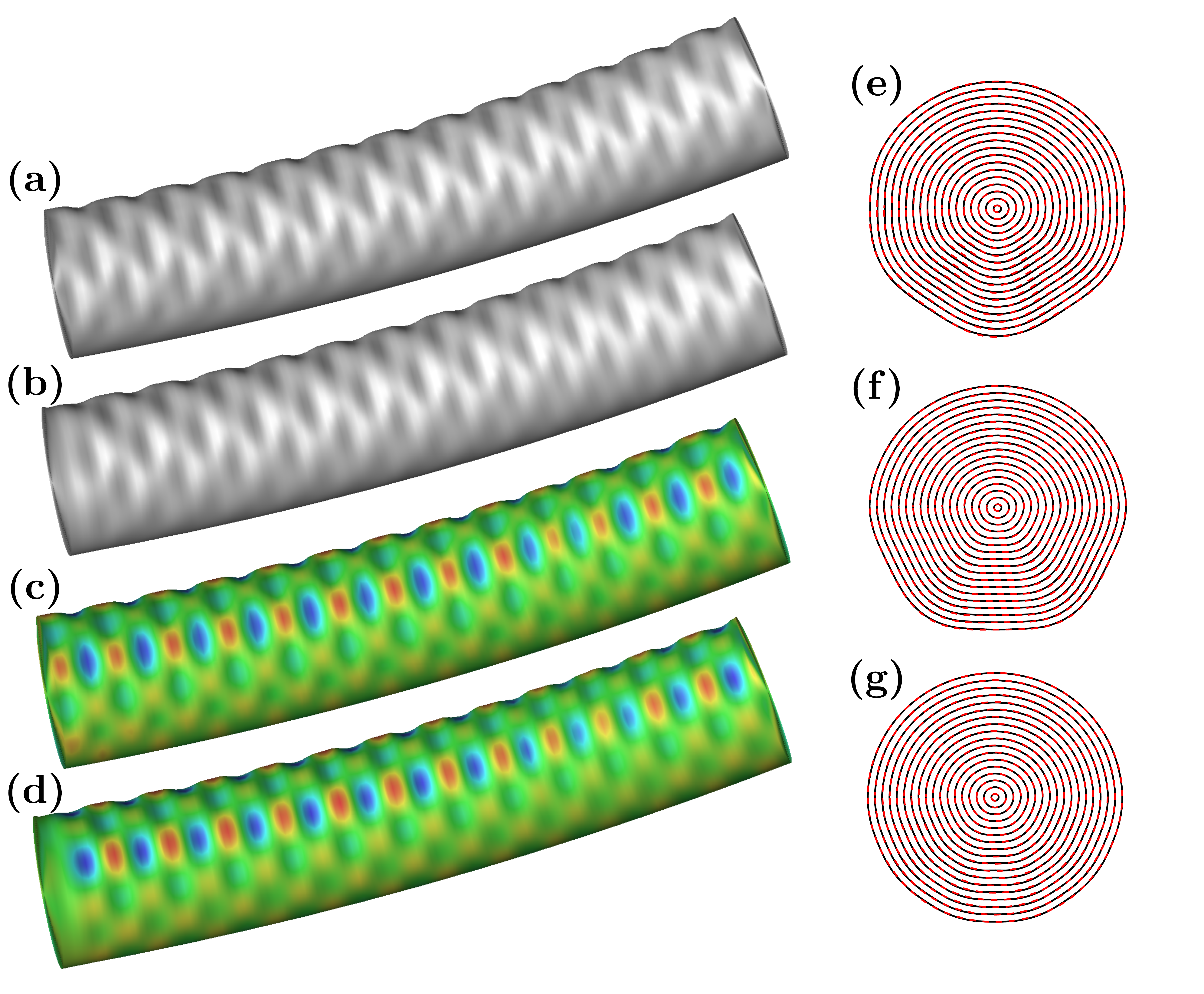}
    \caption{Bent 35--walled CNT obtained via AC (a,c)  and  DML (b,d) model.(c) and (d) show colormap of radial deformations corresponding to (a) and (b). (e--g) Alternate walls of cross--sections obtained via  AC ({\black{$\boldsymbol{-}$}}) and DML ({\red{$\boldsymbol{- -}$}}) model.}  
    \label{fig:bending_accuracy}
\end{figure}

While predicting through the DML model, for a given MWCNT system and loading, at first, the DNN predicts the CoFPCs, which lie in the low--dimensional \emph{latent space}. Subsequently, the high--dimensional deformed 
\black{cross-sections containing all the walls (fig }\ref{fig:torsion_accuracy}\black{ (bottom) and fig }\ref{fig:bending_accuracy}\black{ e, f, g)} is obtained  through 
\black{inverse c--FPCA}. \black{Further, these deformed cross--sections are concatenated  through the length coordinate to generate the 3D deformed shape.} 
Since the functional--PCs are  non--zero almost everywhere, it is imperative that we predict CoFPCs very accurately. 
To achieve very high accuracy for DNNs we have   adopted the following strategies: (i) regularization techniques, (ii)  hyper--parameter tuning, and (iii) features--normalization, (see 
\black{Sec.}~\ref{sec:DNN}). 

The high accuracy of the DNNs is demonstrated  through very low relative--mean squared error (order of $10^{-4}$) for the validation data and excellent correlations ($R>0.993$) for the test data as shown in Fig.~\ref{fig:coorelation_torsion_bending}.

\begin{figure}[h]
    \centering
        \includegraphics[width=1\linewidth]{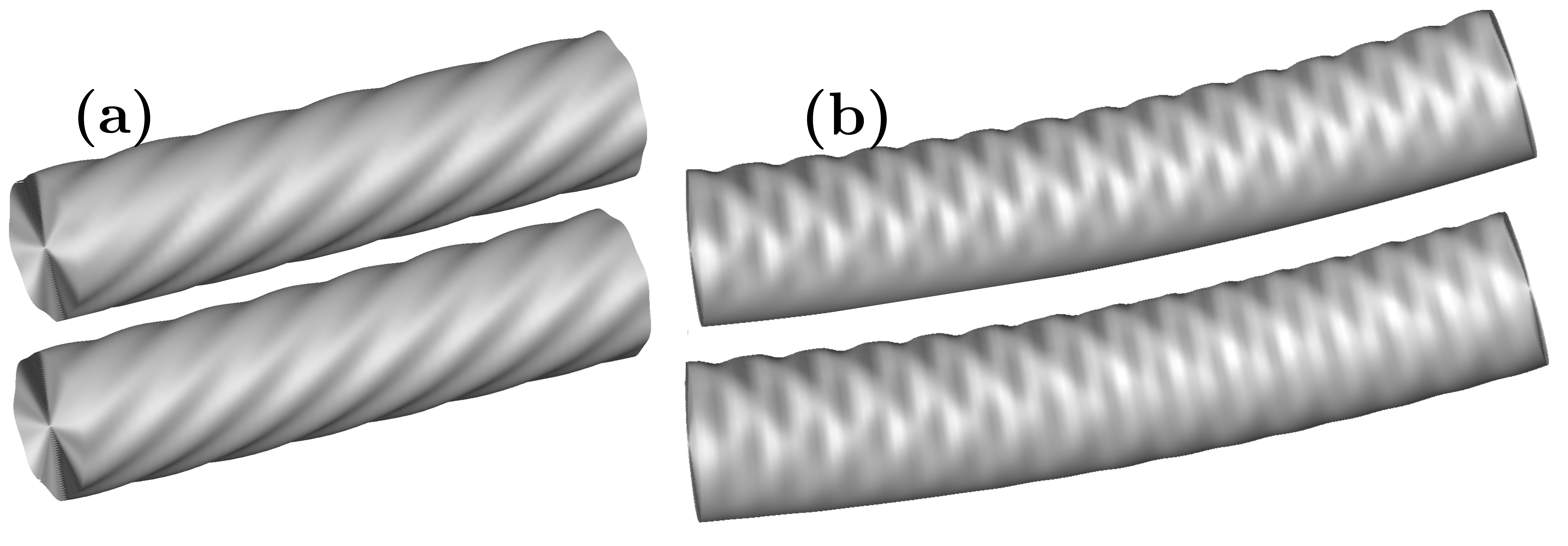}
    \caption{Comparison of  AC (top) and DML (bottom) models for a 32 walled CNT (system which is not a part of training data) under torsion (a) and bending (b).}
    \label{fig:unknown_torsion_32_walled}
\end{figure}

Predictions by the proposed DML model is compared against the
AC model for two types of systems: (i) known systems but unknown loading, 
(ii)  unknown systems and unknown loading, 
Deformation morphologies under torsion and bending obtained through AC and DML models are provided for the known and unknown systems in Fig.~(\ref{fig:torsion_accuracy},\ref{fig:bending_accuracy}) and in  Fig.~\ref{fig:unknown_torsion_32_walled} respectively. The proposed DML model matches remarkably well with the AC model for unknown loading as evident from the deformed surfaces and cross--sections. Their match is quite accurate even when both the system and the loading are unknown \black{(as long as the unknown system is within the range of the training data)}. This obviates the need for 
AC simulations for such systems, yielding huge computational savings.
\black{However, if an MWCNT lies way  outside the range of the training data its accuracy might go down since it might exhibit a  deformation pattern that doesn't occur in the training.} 

\begin{figure}[h]
    \centering
    \includegraphics[width=1\linewidth]{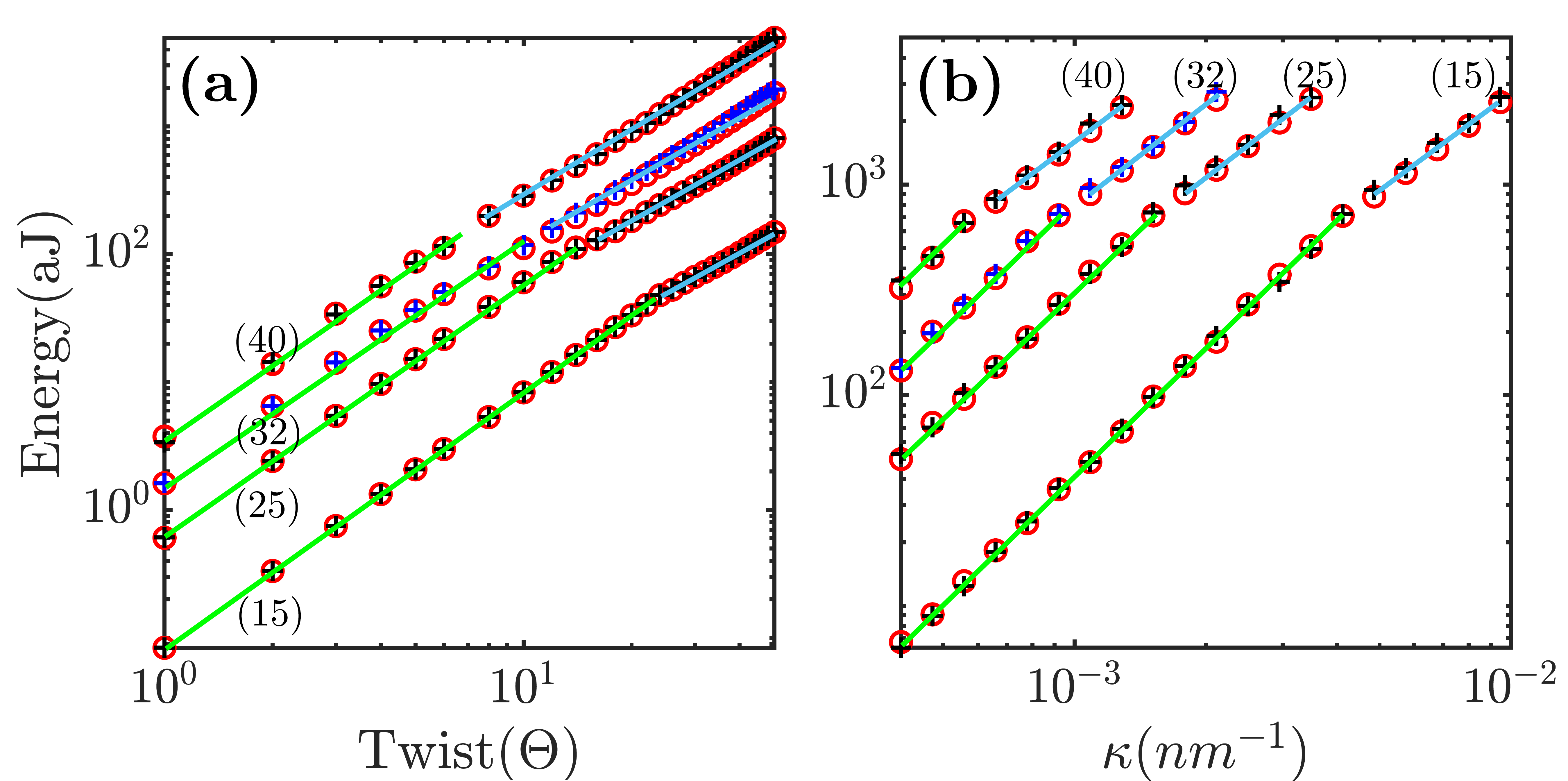}
    \caption{Energy comparison for (15, 25, 32, 40 --walled)  MWCNTs under torsion (a)  and bending (b) via AC (\red{o}) and DML model (\texttt{+}). 32 walled CNT ($\color{blue}{\texttt{+}}$) is an unknown system. The lines ({\green{$\boldsymbol{-}$}}) and ({\postb{$\boldsymbol{-}$}}) are drawn to highlight pre-- and post--buckling regimes.}

    \label{fig:energy_comp}
\end{figure}

The maximum relative error is found to be $\approx$ 1\% for the 32--walled CNT, which is an unknown system under unknown loading.

\black{The deformation obtained from CoFPCs (output of the DML model) is used to compute the energy via a  discretization.} The energy computed through the DML and the AC model matches very well for both known and unknown systems, as shown in  Fig.~\ref{fig:energy_comp}.
The proposed model is significantly  more  efficient than the AC model. The AC model requires tens or hundreds of \emph{total CPU} hours in parallel processing to simulate each of the MWCNTs.
Whereas, inference via the proposed model (upon training), requires only about ten seconds for an unknown MWCNT. 

\begin{figure}[h]
    \centering
    \includegraphics[width=1\linewidth]{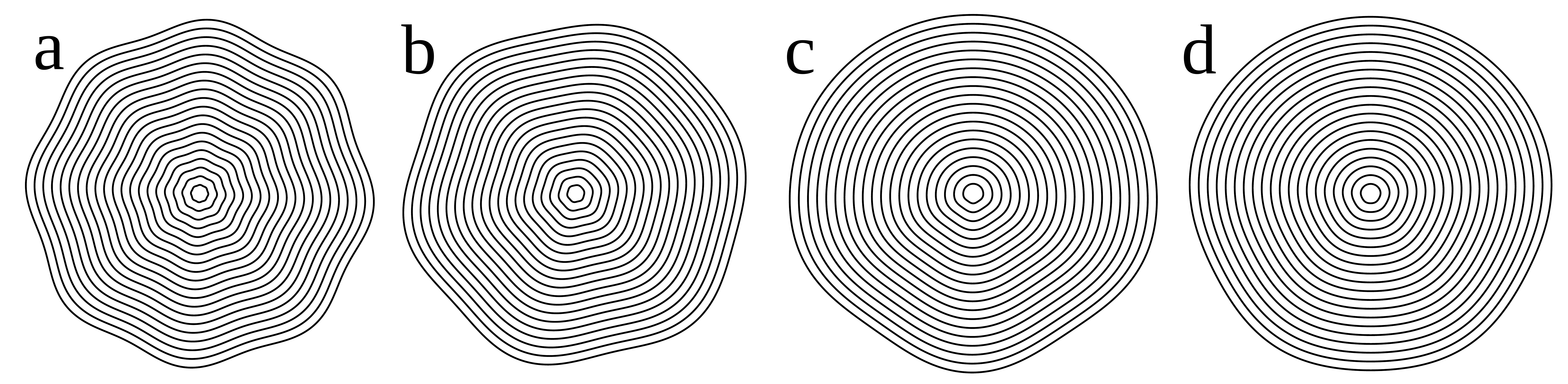}
    \caption{Functional principal components of MWCNTs under torsion (a,b) and bending (c,d).}
    \label{fig:fPCA_components}
\end{figure}

\begin{subsection}{Interpretability of the model}

Herein, we explore the \black{model--based--}interpretability \black{(as defined in }\cite{Murdoch2019}) of the proposed model. 
The proposed model can extract dominant (principal) modes of deformed configurations and their relative contribution in an unsupervised manner. 
A few principal components of the deformation of MWCNTs under torsion and bending are shown in Fig.~\ref{fig:fPCA_components}.  
The rippling deformation of MWCNTs under torsion follows a sequence of ridge and furrows, whereas, in case of bending it resembles the diamond buckling pattern \cite{Arias2008PRL}. 
These key patterns of deformation are captured through the functional principal components (Fig.~\ref{fig:fPCA_components}).
So far these key deformed patterns were  approximately--identified manually for individual MWCNTs 
The principal components of deformation automatically identified in the present model show qualitative similarity with those  identified manually in \cite{Arroyo2003PRL,Arroyo2004IJNME,zou2009effective}. 
These functional--PCs are universal since they are obtained from the entire data set. This fact enhances the model's predictive capability on unseen systems and hence explains the generalizability (performance for  unseen systems) of the model. 
The DNNs learn the reduced dimension spanned by the functional--PCs. The principal modes of deformations are easy to comprehend thus enhance the understanding of how the proposed model works, which makes it an interpretable model. 
\end{subsection}

\section{Conclusion and Discussions} 
In this study, a novel interpretable machine learning model is proposed, which predicts high--dimensional deformed configurations of MWCNTs accurately and efficiently \black{using only 4 inputs}. 

To conclude this study, we summarize its main features. 
\black{\emph{Firstly}, a novel dimensionality reduction technique is proposed that extends FPCA to respect the constraints of deformation exactly. This improves accuracy in low--dimensional representation of deformation and enables accurate  prediction of high--dimensional deformation of MWCNTs.}   
\emph{Secondly}, the proposed model is remarkably accurate for unknown systems and unknown loading . This capability eliminates expensive 
AC simulations for systems beyond what is used in the training, yielding a massive  gain in computational efficiency.
\emph{Thirdly}, the principal components are comprehensible and thus help to elucidate how the model predicts high--dimensional deformation through learning the space of functional--PCs, leading to model--interpretability.

\textbf{Acknowledgments:} 
We acknowledge NSF (CMMI MoMS) grant number 1937983, and HPC facilty SUPERIOR at MTU and XSEDE (Request \# MSS200004). 

\bibliography{main}

\end{document}